\documentclass[sigconf]{acmart}

\usepackage{amsmath}
\usepackage{amsthm,amsfonts,amssymb}
\usepackage{bm}

\usepackage{fullpage,microtype,indentfirst}
\usepackage{fancyhdr}

\usepackage{float}
\floatstyle{plaintop}
\restylefloat{table}

\usepackage{graphicx,subcaption}
\usepackage{algorithm}
\usepackage{algpseudocode}
\usepackage{enumitem}

\usepackage{color}

\usepackage{hyperref}

\newcommand{\N}{\mbox{\rm \hbox{I\kern-.15em\hbox{N}}}}
\newcommand{\R}{\mbox{\rm \hbox{I\kern-.15em\hbox{R}}}}

\def \N {\mbox{\rm \hbox{I\kern-.15em\hbox{N}}}}
\def \R {\mbox{\rm \hbox{I\kern-.15em\hbox{R}}}}

\newcommand{\bA}{\mathbf{A}}

\newcommand{\bC}{\mathbf{C}}

\newcommand{\bF}{\mathbf{F}}

\newcommand{\bI}{\mathbf{I}}
\newcommand{\bJ}{\mathbf{J}}

\newcommand{\bM}{\mathbf{M}}
\newcommand{\bR}{\mathbf{R}}

\newcommand{\bn}{\mathbf{n}}
\newcommand{\bp}{\mathbf{p}}

\newcommand{\bt}{\mathbf{t}}

\newcommand{\bx}{\mathbf{x}}

\newcommand{\bq}{\mathbf{q}}

\begin{document}
\title{Metamorphs: Bistable Planar Structures}
%
\author{Gaurav Bharaj}
\affiliation{%
  \institution{Harvard University and Adobe Inc.}
  }
\email{bharaj@g.harvard.edu}
\author{Danny Kaufman}
\affiliation{%
  \institution{Adobe Inc.}
}
\author{Etienne Vouga}
\affiliation{
 \institution{University of Texas, Austin}
 }
\author{Hanspeter Pfister}
\affiliation{
 \institution{Harvard University}
 }
%
\renewcommand\shortauthors{Bharaj, G. et al}
\begin{teaserfigure}
  \centering
  \includegraphics[width=\textwidth]{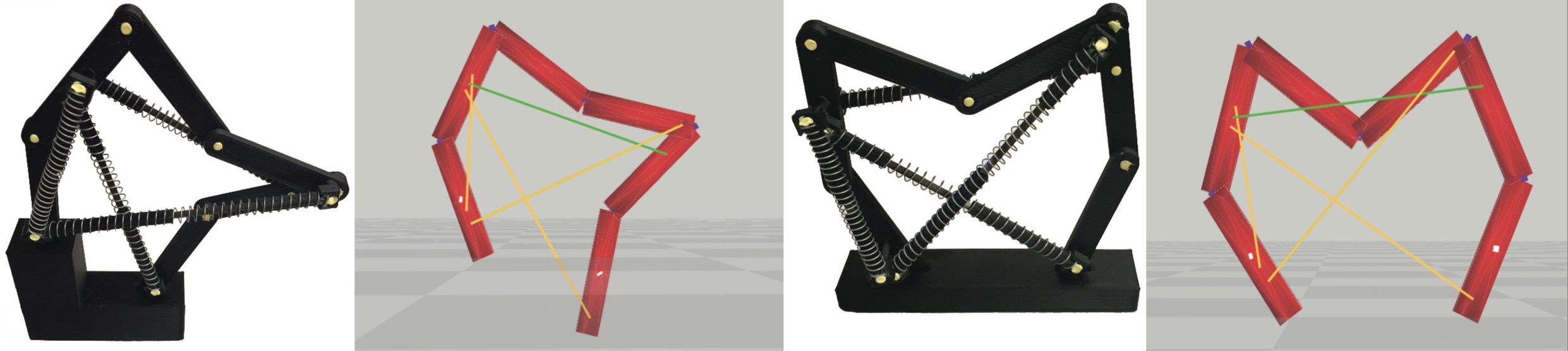}
  \caption[Bistable structures generated by our system]{Bistable structures generated by our system. Optimized and simulated structure with two forms -- duck and teddy. The red bars represent the forms of the structure, while the yellow and green lines represent the springs. The corresponding fabricated results are shown with black bars and metallic springs.}
  \label{fig:metateaser}
\end{teaserfigure}
\begin{abstract}
Extreme deformation can drastically morph a structure from one structural form into another. Programming such deformation properties into the structure is often challenging and in many cases an impossible task. The morphed forms do not hold and usually relapse to the original form, where the structure is in its lowest energy state. For example, a stick, when bent, resists its bent form and tends to go back to its initial straight form, where it holds the least amount of potential energy. 
\\
In this project, we present a computational design method which can create fabricable planar structure that can morph into two different bistable forms. Once the user provides the initial desired forms, the method automatically creates support structures (internal springs), such that, the structure can not only morph, but also hold the respective forms under external force application. We achieve this through an iterative nonlinear optimization strategy for shaping the potential energy of the structure in the two forms simultaneously. Our approach guarantees first and second-order stability with respect to the potential energy of the bistable structure.
\end{abstract}
%
%
%
%
%
\keywords{Computation Fabrication, Physics-based Animation}
\maketitle
%
\section{Introduction}
\begin{figure*}
    \includegraphics[width=\linewidth]{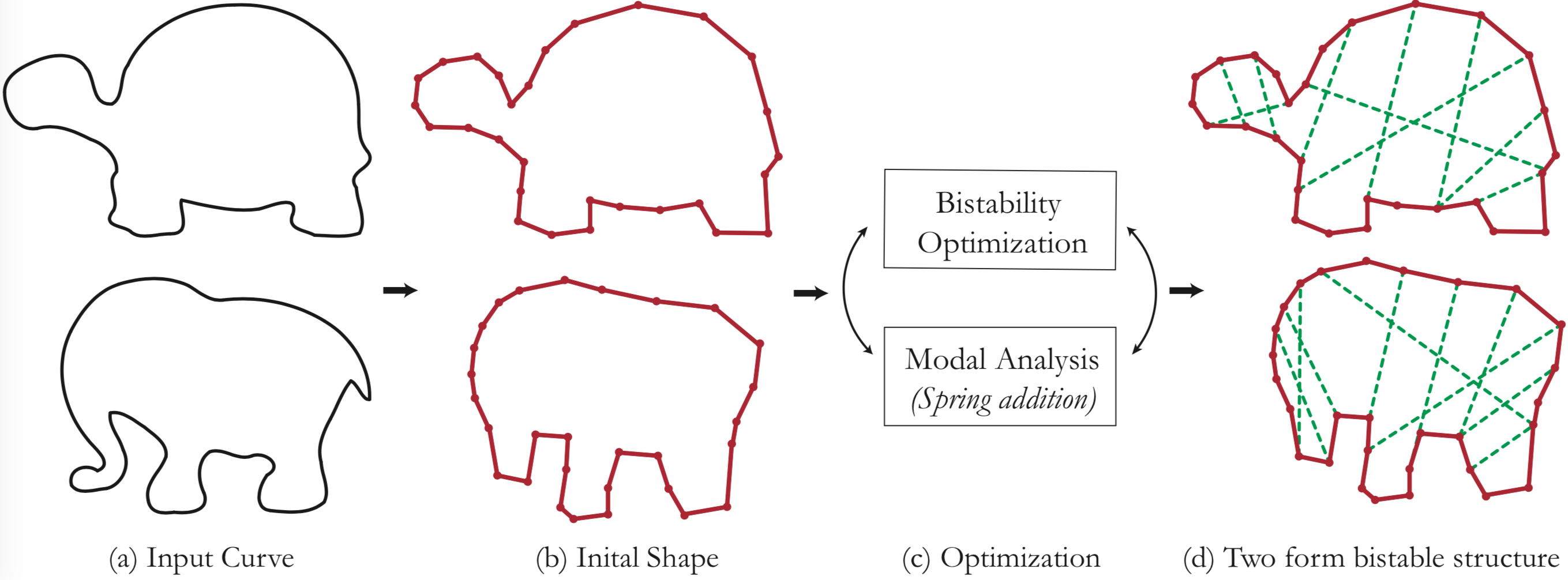}
    \caption[Bistable structures: Overview]{Overview: (a) The input curves of the proposed structures (b) Equivalent planar linkage structures, (c) iterative nonlinear optimization, (d) optimized bistable structure}
    \label{fig:metaoverview}
\end{figure*}
Controlling the morphing and stability properties of a structure under varied force application has been an active area of research in continuum mechanics, robotics and graphics communities. While in continuum mechanics, these deformations are often used to to create functional and compliant objects, in robotics, morphed forms are used to create mechanisms, including soft-robots for safer human interaction applications. These morphed forms often have small deformations as compared to size of the initial form structure and do not have the stability guarantee when morphed. In computer graphics the emphasis is to create artist tools for extreme virtual deformations. While these methods work fluently for animators, they can not be employed for bistable \emph{morphable} structures once fabricated. Other methods are limited to creating structures with a single stable form under external force applications e.g. gravitational forces.

Thus, an interesting question arises: Are there methods for creating two statically stable, morphable, and fabricable structural forms? One way to achieve extremely different forms for a single structure is deform it until it buckles and drastically changes its form. This notion of buckling of metallic beams has existed in the field of continuum mechanics for several decades, where the emphasis had been to avoid buckling of metallic structures. At this point of buckling the structure would permanently deforms or damages. \cite{Bertoldi:2010:NPR} showed the use of the buckling principle for extreme structural deformation of elastic structures. While there are research works where extreme deformation is exploited to create functional objects such as \cite{Yang:2015:Buckling},\cite{Overvelde:2016:OITM}, however, there are no computational methods for creating example-based morphable bistable structures under external forces.

The goal of this work is to create a structure with two different statically stable forms, we call these structures \textbf{\emph{Metamorphs}}, due to their morphable properties. A simple hinged linkage structures with embedded springs is introduced. This hybrid hard-and-soft \emph{springy-linkage} can undergo extreme deformations to morph into different forms. We achieve the stability guarantees by optimizing for first and second order stability of the structure's potential energy, as explained in Section \ref{sec:metaoverview}.

The proposed method can not only be used to create structure with varied forms, but also in the field of soft robotics for creating safe grippers, that can grip objects with various non-convex shapes, while not being specifically programmed for any particular shape. Other applications include creating bistable wing configuration of aeroplanes \cite{Thill:2008:MS}. Such plane's wings can adapt to various turbulence  conditions with increased efficiency or use different wing structural forms while take-off, landing and cruising. Finally, similar to satellite wings, that are packed according to origami principles \cite{Miura:1985:MPD}, Metamorphs can be used in applications where a different packed and unpacked stable forms are needed.
\\
Metamorphs also find use in animatroics. Artists create puppets and articulated character where a deforming structure can be used to express the various \emph{moods} of a character, and in story-telling mediums such as pop-up books \cite{Li:2010:POPUP} by creating collapsible structure structures. Lastly, such a method may also be used to create shape shifting furniture and human-computer interaction devices \cite{Yao:2013:PPA}. A single piece of Metamorph furniture can be configured to take different functional forms, or folded into a space-saving form. For example, a structure can be used as a table \emph{or} morphed into stool \emph{or} morphed into a compact form.
\section{Overview}\label{sec:metaoverview}
This sections introduces the notion of stability and provides a general overview of our approach. Figure \ref{fig:metaoverview} summarizes the same.

\paragraph{Structure and kinematics}
The input to the method are two input curves used to create the forms that a linkage structure must morph into. We define our structures as rigid bars that are connected at designated end-points via hinge-joints. All forms of a structure share the same bar count and connectivity and are geometrically equivalent. Details on how the linkage-based shapes are created from input curves can be found Section \ref{sec:kinematics}. The user can also fix certain linkages as fixed if desired. Between the various bars, springs are added, these spring are used for to create stability for the two forms of the structure. For the purposes of this work, all the bars and springs are planar while having a certain \(z\)-depth, hence the problem essentially simplifies to 2D. This simplification is done in-order to create a fabricable structure. Adding springs that lie on different planes in 3D would lead to self-intersections, and unfit for fabrication.

\begin{figure}
\centering
\includegraphics[width=\linewidth]{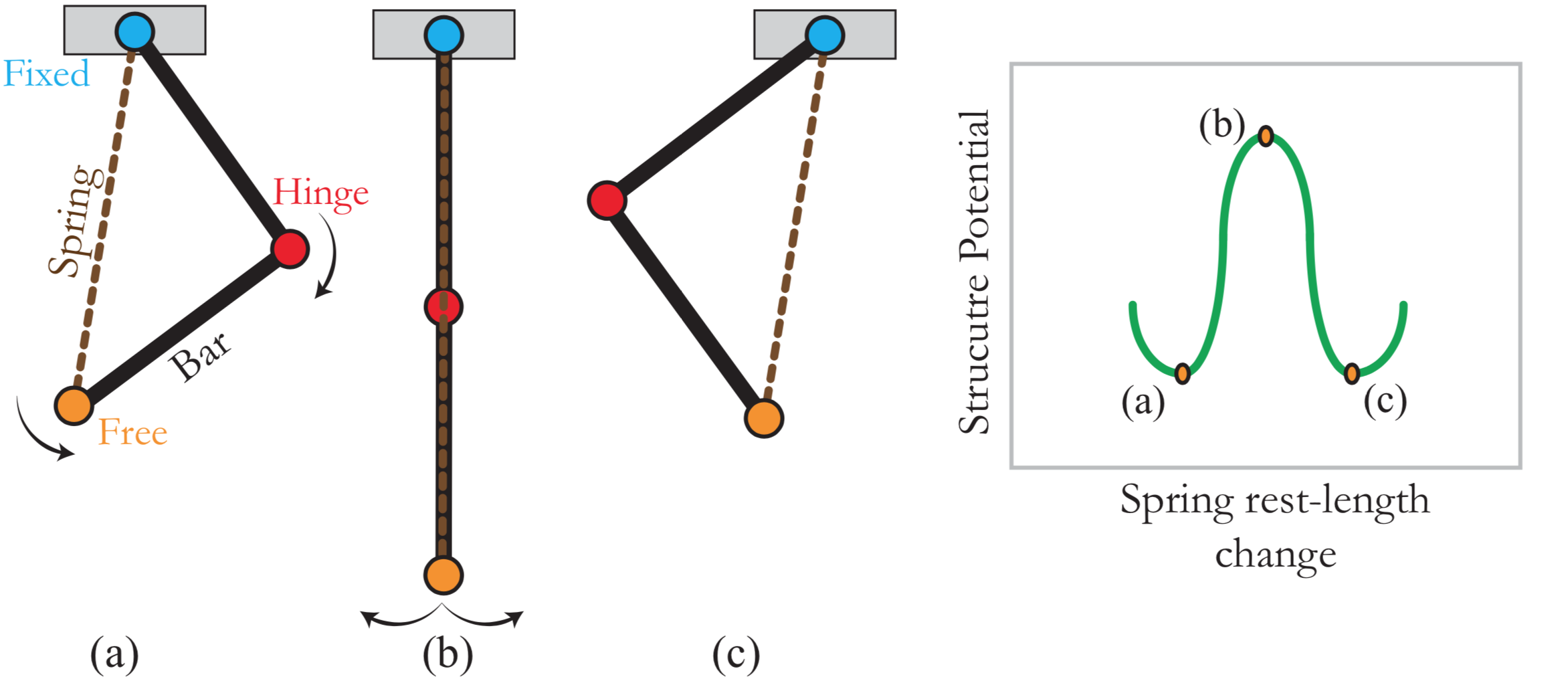}
\caption[Springy Linkage]{Left: Springy linkage in various stable forms. Right: Corresponding plot of spring's length vs. structure's potential energy}
\label{fig:springylinkage}
\end{figure}

\paragraph{Energy-based Stability}
\label{sec:energyStability}
Consider two bars connected at a hinge. Let the upper bar be fixed at the outer end. Next, we add a spring connecting the outer ends of the two bars. For a fixed rest-length of the spring, the \emph{springy linkage} can take a particular kinematic form as shown in Figure \ref{fig:springylinkage} (a). The figure on the far right shows the plot of change in spring's rest-length vs. the potential energy gained by the structure as a result of change in the spring's length. When the spring is stretched the most, form (b), the springy linkage has the largest potential energy. At this stage (i.e. the point of bifurcation) the system can morph back into form (a) and with an equal probability morph into form (c), the second state at that the structure has the lowest potential energy. This phenomenon is also called \textbf{\emph{bistability}}. The point of inflection in form (b) is also called point of bifurcation or buckling.

\paragraph{First-Order Energy Stability}
Similar to the \emph{First-order necessary conditions for optimality}, \cite{Wright:1999:NO}, first-order stability of potential energy \(\mathrm{V}(\mathbf{x})\) is the state of the structure where the gradient of the energy potential is zero, i.e. \(\nabla_{\mathbf{x}}\mathrm{V}(\mathbf{x})=0\). At this point, the rate of change of the energy in any direction (locally) is zero. Since rate of change of the potential represents the forces acting on the system, the first-order stability intuitively means that forces acting on the system balance out and total forces acting on the system are zero. For example, in the case above, at all three forms (a), (b) and (c) the forces balance out. However, in form (b), the system is not truly stable, as even a infinitesimal push will lead to the structure morphing into forms (a) or (c). This morph depends on the direction of the push (force or torque). Hence, first-order stability is a \emph{necessary}, but not \emph{sufficient} condition for establishing stability.

\paragraph{Second-Order Energy Stability}
If we want a truly stable structure, the it has to be a second-order stable structure. Similar to the \emph{Second-Order Optimality Conditions} used in convex optimization, this condition suggests that second-order derivative of the energy should be greater than zero. For a \(n\)-dimensional structure, this suggests that the Hessian \(\mathcal{H}\) of the energy potential should be positive definite. That is, \(\mathcal{H}(\mathbf{x})= \nabla^2_{\mathbf{x}}\mathrm{V}(\mathbf{x})\succeq 0\) where \(\mathbf{x}\in\mathbf{R}^n\). In the example shown in Figure \ref{fig:springylinkage}, while (a) and (c) satisfy the second-order stability condition, (b) does not. Thus by optimizing for first and second-order stability simultaneously, we can create a morphable structures with two stable forms.

To summarize, we start with a kinematically feasible initial structure (Section~\ref{sec:kinematics}). That is, the linkage structures can morph into the two desired forms, while they may not achieve stability or retain these forms. These initial forms can be chains, or branched, or loopy structures. An example of the initial design is shown in Figure~\ref{fig:input}.

In order to model the physical energy of the of linkage structure and joints, we then introduce in Section~\ref{sec:physicalmodeling} a rigid-body framework. Since our problem is modeling static stability, we derive the system equations, Euler-Lagrange equations~\cite{goldstein2011classical}, such that the velocity terms are zero, that simplifies our formulation. We also introduce an iterative springs addition algorithm (Algorithm~\ref{sec:itr}) to create statically stable structural forms. In Section~\ref{sec:metaoptimization}, the notion of energy shaping for creating stable structures is detailed. The method guarantees second-order stability w.r.t energy potential, there by making sure that the structural forms are stable, and robust against gravity and user employed forces.

Section~\ref{sec:metaresults} discusses the results. We fabricated some examples for validation, and show complex examples virtually. We also show examples with real world functional applications, for example, shape shifting wings. Finally, Section~\ref{sec:future} presents the conclusion and a discussion of the limitations, and opportunities for future work. To summarize, our computational design method introduces the notion of second-order static stability for bistable structures with the following major contributions:
\begin{enumerate}
    \item Novel computation design tool for morphable structures.
    \item Novel optimization formulation for creating bistable structure forms via energy shaping.
    \item Novel spring assembly process for Hook's springs.
\end{enumerate}
\section{Related-Works}
Creating controllable deformation has been an active area of research among various interdisciplinary areas. We now discuss some of the state-of-the-art techniques that exist.

\paragraph{Geometric and kinematic design:}
As the name suggests, methods defined in this category use the geometric shape and kinematics (motion) to define the numerical simulation of the design problem. Here, the physical energy of the system is \emph{not} modeled, but the shape and motion are defined through complex mathematical functions and optimization. For example, in linkage design problems, \cite{Coros:2013:CDM} define the movement of the linkage by a constraint optimization of the connections; however, they do not model or optimize for the physical energy of the linkage structure. Some methods in this category aim to bring virtual characters to the real world. It is now possible to create 3D printable representations of virtual linkage-based characters with joints~\cite{Cali:2012:NAM:2366145.2366149}, and mechanical toys capable of interesting non-walking motions~\cite{Ceylan2013,Thomaszewski14Linkages,yu2017computational}. Origami inspired geometric design also falls in this categories, where rigid origami \cite{o1998folding,dudte2016programming}, and pop-design \cite{Li:2010:POPUP} are used to create kinematics of a design. \cite{Mitani:2004:MPT} show the use of strip patterns to assemble 3D models. Other researchers like \cite{Skouras:2015:ISD}, create complex structures by creating user interfaces for interlocking elements by understanding the geometry of the atomic-elements, while \cite{hildebrand2012crdbrd} create puzzles by using planar slits abstracted from 3D shapes. \cite{xin-2011-making} create 3D puzzles by automatically disintegrating 3D models into interlocking elements.

\paragraph{Physical energy-based design}
In these works, the aim is to optimize the material and shape properties, where cost functions model the efficiency of functionality through \emph{physically-based} numerical simulation. Such works use principles from continuum mechanics (finite element method, boundary element methods, etc), fluid mechanics, and physics-based wave propagation (optics) to model the material deformation and response and rigid body dynamics to model the energy dynamics and ground contacts. Various problems that have been worked on include: 

Appearance-based material distribution for subsurface scattering~\cite{Hasan:2010:PRO,Dong:2010:FSS}, caustics~\cite{Marios:2011} or reflectivity~\cite{Matusik:2009,Weyrich:2009:FMF}, material and physical \cite{Bickel:2010:DFM} behavior of fabricable shapes. \cite{Prevost:2013:MIS} and \cite{Musialski:2015:RSO} optimize shape via material carving to control the moment-of-inertia property of the rigid shapes, and as a result control static stability, given that the shape rests on a surface or on water. \cite{Li:2016:acoustic_voxels} optimize for the sound spectrum through voxel filters that act much like selective damping filters, while \cite{Martin:2015:ODO} use fluid-dynamics principles to model the uplift and drag for 3D flying designs. Creating controllable deformation has been an active area of research among various interdisciplinary areas. \cite{Skouras:2013:CDAD}, \cite{Schumacher:2015:MCE}, \cite{bern2017interactive} and \cite{perez2017computational} optimize for shape deformation properties to create articulate characters and shapes. \cite{Chen:2017:DNC} created jumping robots with precise upright landing capabilities, by modeling the dynamics of robots. Finally, \cite{Chen:2013:SRM} abstract previous methods by goal, parameter reduction scheme, optimization method, and simulation algorithm and provide a structured way to define computational design problems.
%
%
\\
\\
Below we discuss related works in several interdisciplinary areas including works from computer graphics and animation, robotics, and continuum mechanics communities.

\paragraph{Virtual Deformation Design:}
Creating user-controllable deformation to create an articulate virtual character has seen many wonderful research contributions lately. Here, the developed methods are user-assisted, and semi-automatic for physically plausible articulation. \cite{Martin:2011:EEM} and the references within create virtual example-based material deformations, where the user-provided shape forms are replicated under force application. These method are created for physical plausibility, and cannot be applied for fabrication. Along similar lines, \cite{Coros:2012:DOA} present a method for creating virtual deformable characters with toon-like articulation, where the secondary animations are automatically created. While \cite{Xu:2015:IMD} create deformation models for elasticity of continuum's material, and recently \cite{Xu:2017:EBD} provide a method for controlling the damping behavior for materials undergoing deformation. Such works are differentiated from the task of rigging-based (skinning) deformation \cite{bharaj2012automatically}, \cite{Baran:2007:ARA}, where the deformation is based on non-physical deformation energy.

\paragraph{Extreme Mechanics:}
Extreme mechanics is a sub-field of continuum mechanics, where large deformations of elastic materials and shapes are explored for controllable deformation. \cite{Bertoldi:2010:NPR} create negative Poisson-ratio structures (which expand when compressed) by using the buckling principle of deformation. \cite{Yang:2015:Buckling} take this further and create movements such as rotation, extension, etc for pneumatically actuated soft-robots. These methods however, do not provide a design tool for creating example-based extreme deformations and are limited to the premeditated deformation types. Greater emphasis has been laid on understanding the real life properties of these deformations: for example, \cite{Marchese:2016:DTO} use vision-based data-driven methods to create soft gripper, where system equations are learnt for predictable deformations. Complete understanding of extremely soft and complaint deformable materials is still a research question and hence using soft materials with nonlinear deformation behaviors can prove tedious and unpredictable especially for large deformations. This led us to the use Hook's springs for extreme deformation.

\paragraph{Mechanism Design:}
\cite{Coros:2013:CDMC} developed an algorithm for design of target curve-based linkage character, while we extend these to create walkabale and statically stable robots. \cite{Zheng:2016:DLC} use scissor linkages for creating shape-shifting characters. They do not optimize for static stability; however, the optimized linkage-based shapes can be packed and unpacked without collisions. \cite{Gauge:2015:IDM} similarly create characters connected by elastic wires.
\cite{Culpepper:2004:CM} create a simple shape shifter, where the emphasis is on the design of the mechanism for bistable shapes. \cite{Ou:2016:AHI} create pneumatically actuated shape shifters. In these works, the amount of deformation remains small, and the overall forms of the shape remain the same. \cite{Ion:2017:DMM} use a bistable mechanics primitive to create programmable logic gates such as AND, OR, XOR, and show how these simple \emph{physical} computing logics can be exploited. 

\paragraph{Stability Optimization:}
Most computational design methods strive to create controllable shapes. The notation of static stability is to create an object that is true to its shape under external forces and perturbations. The notion of stability comes up in many forms over an array of research works. \cite{Chen14:ANM} optimize material properties for creating fabricable stable structures that have been optimized to \emph{hold a single shape} under gravity or preset forces. \cite{Garg:2014:WMD} create wire-mesh designs,  with the notion of a stable shape under gravity; similarly, \cite{Zehnder:2016:DSO} and \cite{Miguel:2016:CDS} create intricate object designs such that objects can hold their forms for a single form. Moreover, there is no notion of bistable structures or morphing. \cite{Panetta:2015:ETAF} create deformable structures that deform under constant external force loads (other than gravity) to create different shapes. The deformations are small and not statically stable without the constant external force loads. 
\section{Problem Formulation}
\subsection{Structure from input curves}\label{sec:kinematics}
\begin{figure}
    \centering
    \includegraphics[width=\linewidth]{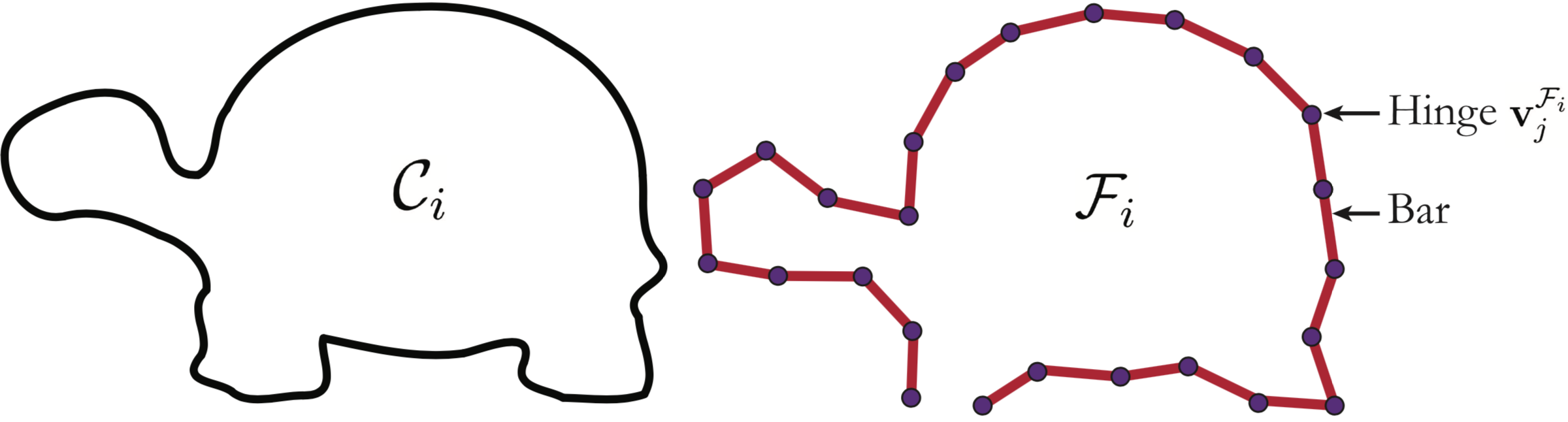}
    \caption[The input curves]{Left: The input curve, $\mathcal{C}_i$. Right: The calculated planar structural form, $\mathcal{F}_i$ (in red) as explained in Section \ref{sec:rigidDeformations}}
    \label{fig:input}
\end{figure}

Input to the method are two 2D curves, such as b\'ezier curves, these curves define the two forms for the proposed bistable structure. There are no restriction on convexity or continuity of curves. Both the curves are spatially normalized and translate so the center of mass lies at the origin. Each form is defined using $m$ rigid bars. For each curve corresponding $m$ points on the curve are defined which serve as the correspondence points for an end of the $m$ bars, with a prescribed bar length. Alternatively, the form curves are sampled into $m$-points which serve as corresponding input bar ends. Let us call the input form curves as $\mathcal{C}_i$ and the linkage structure forms as $\mathcal{F}_i$ for the $i^{th}$ form. It it also assumed that each bar of $\mathcal{F}_i$ is connected to its immediate neighbor by a hinge connection (Section \ref{sec:physicalmodeling}). We now define an optimization algorithm that is used to create form $\mathcal{F}_i$ which is as-close-as-possible to $\mathcal{C}_i$, while keeping the rigid bar assumption.
\paragraph{Rigid Deformation}
\label{sec:rigidDeformations}
A natural disposition is to form an deformation energy as described by \cite{Sorkine:2007:ARAP}. However, this method does not guarantee rigid-deformation, but only an approximation. Hence, we propose the following method.
\\
The input structure form $\mathcal{F}_i$ is be deformed to match $\mathcal{C}_i$ while maintaining rigidity and hinge connectivity. That is, the degrees of freedom of deformation are the rotations of the bars of $\mathcal{F}_i$. One way to formulate this rigid-deformation energy is as follows. Let $\mathbf{v}^{\mathcal{F}_i}_j$ be the $j^{th}$ bar end on $\mathcal{F}_i$ and $\mathbf{v}^{\mathcal{C}_i}_j$ corresponding sample on $\mathcal{C}_i$, as shown in Figure \ref{fig:input}. We want to minimize the distance between these two points while making sure the the neighboring linkages (due to the hinges), $\{\mathbf{v}^{\mathcal{F}_i}_{j+1}, \mathbf{v}^{\mathcal{F}_i}_{j-1}\}$ are at a prescribed distance. This can be formulated as the following optimization:
\begin{align}
\arg\min_{\{\mathbf{v}^{\mathcal{F}_i}_j\}} \frac{1}{2} |\mathbf{v}^{\mathcal{F}_i}_j-\mathbf{v}^{\mathcal{C}_i}_j|_2^2\\
s.t.\;|\mathbf{v}^{\mathcal{F}_i}_j-\mathbf{v}^{\mathcal{F}_i}_{j-1}|^2_2 &= c\label{eq:rd1}\\
|\mathbf{v}^{\mathcal{F}_i}_j-\mathbf{v}^{\mathcal{F}_i}_{j+1}|^2_2 &= c\label{eq:rd2}\\
\forall j \in \{1, 2, ...,\;&m\}\nonumber
\end{align}
Here $|\;.\;|_2^2$ represents the squared L2-norm. The vertex constraints given by equations \ref{eq:rd1} and \ref{eq:rd2} can be written in the matrix view as well. The above is a quadratic cost with quadratic constraints and can solved via a quadratic programming solver with quadratic constraints. We repeat this procedure for both forms of a Metamorph structure, such that each curve is sampled into $m$ samples, and hence all forms have the same geometry and kinematics.
\subsection{Physical Modeling}\label{sec:physicalmodeling}
The numerical simulation of the linkage structures is modeled via rigid-body dynamics equations (Appendix~\ref{sec:simulation}). Two types of constraints used for modeling linkage structures are:
%
\begin{enumerate}
\item \textbf{Hinge/Pin Joints}: Constraints the position and two orthogonal rotational degrees of freedom at local points on two rigid-bodies such that the third rotation
  degree of freedom becomes the hinge-axis along which the bodies can rotate.
\item \textbf{Fixed Joints}: All positional and rotational degrees of freedom are fixed at a certain point on both bodies. The bodies are essentially locked and held fixed at predefined local points.
\end{enumerate}
\paragraph{Constraint Jacobian}
As derived in the Appendix~\ref{sec:simulation} and \cite{Cline:2002:RBD}, let \(\mathcal{C}(\mathbf{q}) = \mathbf{0}\) be the satisfied, that is, let us assume that the kinematic constraints of the structure are always satisfied. Accumulation of all the constraints is represented by the constraint Jacobian \(J\) matrix for the complete rigid-body assembly.
\paragraph{Spring model for a springy linkage}
Given a rigid-body \(\mathcal{B}_i\), the world position \(\mathbf{u}^w_i\) of local point \(\mathbf{u}^l_i\) is given by \(\mathbf{u}^w_i( \mathbf{p}_i, \mathbf{a}_i) = \mathrm{R}_i(\mathbf{a}_i) \mathbf{u}^l_i + \mathbf{p}_i\). Here, \(\mathbf{q}_i = [\mathbf{p}_i\;\;\mathbf{a}_i]^T\) represent the kinematic degree of freedom of rigid-body \(\mathcal{B}_i\). \(\mathbf{p}_i\) is the translational and \(\mathbf{a}_i\) is the axis-angle, the rotational degree-of-freedom of the rigid-body. Then, the spring potential between a local points \(\mathbf{u}^l_1, \mathbf{u}^l_2\) on \(\mathcal{B}_1, \mathcal{B}_2\) respectively is given by:
\begin{align}
\mathrm{V}(\mathbf{x}_{kin}) &= \frac{1}{2} k (l - \sqrt{f(\mathbf{x}_{kin}})\;)^2\\
  \text{where}\;\; \mathbf{x}_{kin} &= \{\mathbf{p}_1, \mathbf{a}_1, \mathbf{p}_2, \mathbf{a}_2\}\\
  \text{and } f(\mathbf{x}) &= g(\mathbf{x})^T\,g(\mathbf{x})\\
  g(\mathbf{x}) &= [\mathrm{R}_1(\mathbf{a}_1) \mathbf{u}^l_1 + \mathbf{p}_1 - \mathrm{R}_2(\mathbf{a}_2) \mathbf{u}^l_2 - \mathbf{p}_2]
\end{align}
Here \(\mathbf{x}_{design} = \{k, l\}\) is the spring stiffness and spring rest-lengths respectively for a spring. The gradients of these quantities with respect to the kinematic degrees of freedom, used to calculate the forces and torques are described in-detail in Appendix \ref{sec:axisangle}.
\section{Optimization}\label{sec:metaoptimization}
\subsection{Iterative spring addition}\label{sec:itr}
We propose the following iterative algorithm for adding springs to create the Metamorphs with stable forms. Starting with the initial forms (without springs), the algorithm automatically calculates where the next spring should be added. The algorithm stops as soon as static stability is achieved for both forms of the structure. As a result the algorithm adds an \emph{optimal} number of springs to the structure which achieves bistability. While the details of each step of the algorithm are described in subsequent sections, Algorithm~\ref{alg:itr} details the outline.
%


%
\begin{algorithm}[t]
\caption{Iterative static stability optimization}
\label{alg:itr}
    \begin{algorithmic}
    \State Calculate $\{\mathcal{F}_i\}$ forms from $\{\mathcal{C}_i\}$ using Rigid Deformation \Comment{Section \ref{sec:rigidDeformations}} \vspace{5px}
    \State \Comment{Let $\{\mathcal{F^*}_i\}$ be the structure with optimal springs and static stability} \vspace{5px}
    \While{(converged == false)}
        \State $b_j$ = \texttt{ModalAnalysis}($\{\mathcal{F}_i\}$)\Comment{Calculate most deformed bar (Section \ref{sec:modal})}\vspace{5px}
        \State \texttt{AddMinEnergySpring}($\{\mathcal{F}_i\}, b_j$)\Comment{Add spring on $b_j$ to all forms (Section \ref{sec:minengsp})}\vspace{5px}
        \State converged = \texttt{StaticStabilityOptimization}($\{\mathcal{F}_i\}$, $\{\mathcal{F^*}_i\}$)
        \State \Comment{Run static-stability optimization (Section \ref{sec:nshapeOpt})}\vspace{5px}
    \EndWhile
    \State $\{\mathcal{F^*}_i\}$ = \texttt{FabricationReform}($\{\mathcal{F^*}_i\}$)\Comment{Spring fabrication reformulation (Section \ref{sec:metafabrefor})}\vspace{5px}
    \end{algorithmic}
\end{algorithm}

\subsection{Reduction}\label{sec:reduction}
Starting from first principles with the force-acceleration rigid-body
dynamics equations along with the added springs forces gives us the
following equation:
\begin{align}
J^T\mathbf{\lambda} + f_{ext} &- \nabla_{\mathbf{x}_{kin}} V(\mathbf{x}_{kin}, \mathbf{x}_{design}) = 0\label{eq:sys}\\
            s.t.\;\;\mathcal{C}(\mathbf{q}) &= \mathbf{0} \text{ (assumed to be true)}\nonumber
\end{align}
Here $f_{ext}$ are external forces such as gravity, while \\
$-\nabla_{\mathbf{x}_{kin}} V(\mathbf{x}_{kin}, \mathbf{x}_{design})$
represents external forces due to spring potential $V$. Here, $\mathcal{J}^T\mathbf{\lambda}$ is due to principle of virtual-work. Equation \ref{eq:sys} above has three kinds of degrees of freedom, namely,
\begin{enumerate}
\item Lagrange multipliers \(\lambda\).
\item \(\mathbf{x}_{kin}\) which are the position and orientation of
  rigid bodies.
\item \(\mathbf{x}_{design}\) which are spring stiffnesses (material parameters) and rest lengths.
\end{enumerate}
In our formulation \(\mathbf{x}_{kin}\) are held fixed, by solving the problem in the null-space \(N_{J^T}\) of the constraint Jacobian, we further reduces the complexity of the optimization. The updated formulation is shown below:
\begin{align}
N_{J^T}(J^T\mathbf{\lambda} + f_{ext} - \nabla_{\mathbf{x}_{kin}} V(\mathbf{x}_{kin}, \mathbf{x}_{design})) &= 0\label{eq:null}\\
N_{J^T}(f_{ext} - \nabla_{\mathbf{x}_{kin}} V(\mathbf{x}_{kin}, \mathbf{x}_{design})) &= 0\label{eq:nullsys}
\end{align}
Equation \ref{eq:null}, leads to Equation \ref{eq:nullsys} as $N_{J^T}J^T = 0$. As a result there is a reduction in the degrees-of-freedom of $|\lambda|$, which is equal to the number of constraints \emph{rows} added by a hinge/pin and fixed joints.
\subsection{First-Order Stability Condition}
For a statically stable system the primary requirement is first-order
stability given by equation \ref{eq:nullsys} above. Thus, we want to satisfy equation \ref{eq:nullsys} for all forms $\mathcal{F}_i$ simultaneous. This equates to the following energy minimization problem:
\begin{align}
\arg\min_{x_{design}} &\frac{1}{2}{|N_{J^T}(f_{ext} - \nabla_{\mathbf{x}_{kin}} V(\mathbf{x}_{kin}, \mathbf{x}_{design}))|_2^2}^{\mathcal{F}_i}\label{eq:fo}\\
\forall &i \in \{1, 2\}\nonumber
\end{align}
Thus, we want to reduce forces (gradient of the energy) acting in the two different forms, which is equivalent to first-order stability for constraint rigid body systems.

\begin{figure}[t]
  \includegraphics[width=\linewidth]{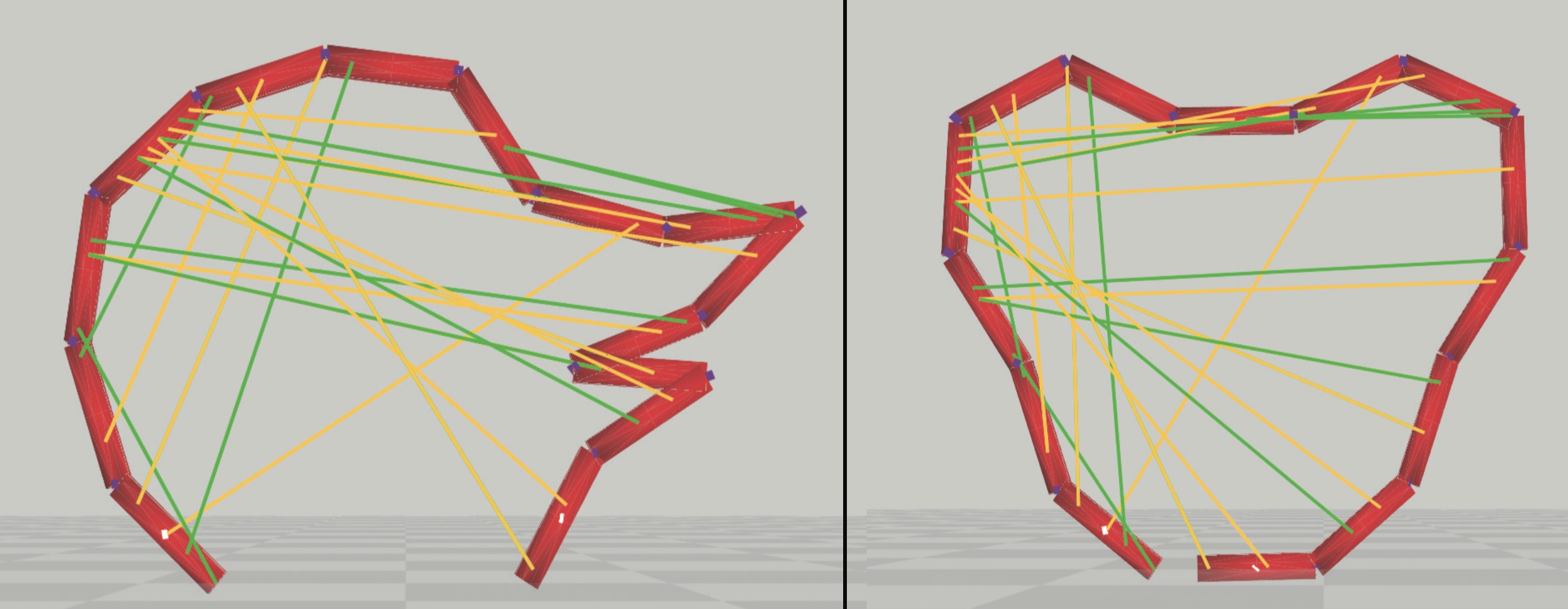}
  \caption[Two morphable forms]{Two morphable forms Duck (left) and Teddy (right)}
  \label{fig:duckTeddy}
\end{figure}
\subsection{Second-Order Stability Condition}

Not only do we want the forces acting on the structure in the two forms $\mathcal{F}_i$ to balance out (equation \ref{eq:nullsys}), but also the structure must guarantee second order stability, such that under local perturbations, the structure returns to its stable forms $\mathcal{F}_i$. This is guaranteed when the hessian of the energy potential is positive-definite as explained in Section \ref{sec:energyStability}. The potential of the spring structure is given by \(V(\mathbf{x}_{kin}, \mathbf{x}_{design})\). In-order to guarantee second-order stability, we want the Hessian \(\mathcal{H}(\mathbf{x}_{design}) = \nabla^2_{\mathbf{x}_{kin}}V(\mathbf{x}_{kin}, \mathbf{x}_{design})\) to be positive-definite.
\\
Ones again we reduce the above to by projecting \(\mathcal{H}\) in the null-space of the constraint-Jacobian, given by \(\mathcal{H}_{N^J} = N_{J^T}^T \mathcal{H} N_{J^T}\). In-order to guarantee that \(\mathcal{H}_{N^J}(\mathbf{x}_{design})\) be positive-definite, we add non-linear constraints of the form \(\mathcal{E}_j(\mathcal{H}_{N^J})> 0\). Here, \(\mathcal{E}_j(\mathcal{H}_{N^J})\) is the \(j^{th}\) eigenvalue of the null projected energy hessian. Thus, for $i^{th}$ form we get constraints of the form:
\begin{align}
{\mathcal{E}_j(\mathcal{H}_{N^J}(\mathbf{x}_{design}))}^{\mathcal{F}_i}> 0\label{eq:so}
\end{align}
\subsection{Minimal Potential Regularizer}
We want to guide the optimization towards a lower energy potentials \(P\) as high energy structures will wound too tight and bound to eventually snap. In case of springs a high potential configuration is the one in which the springs are stretched or compressed much beyond the rest lengths. Although the optimization can balance out the forces and torques caused by such springs, the springed structure can eventually snap. To alleviate this, we add the following regularizer to the optimization energy defined by equation \ref{eq:fo}:
\begin{align}
w (V^{\mathcal{F}_i}) \label{eq:pereg}
\end{align}
Here \(w \in \mathbf{R}^1\) is the potential regularizer.
%
%
\subsection{Two forms optimization}\label{sec:nshapeOpt}
With all the ingredients defined in previous sections, we are now ready to describe the overall optimization strategy for the two forms optimization simultaneously. By combining equations \ref{eq:fo}, \ref{eq:so} and \ref{eq:pereg}, we define the following nonlinear optimization problem:
\begin{align}
\arg\min_{\mathbf{x}_{design}} &\frac{1}{2}{|N_{J^T}(f_{ext} - \nabla_{\mathbf{x}_{kin}} V(\mathbf{x}_{kin}, \mathbf{x}_{design}))|_2^2}^{\mathcal{F}_i}\nonumber\\
&+ w (V(\mathbf{x}_{design})^{\mathcal{F}_i})\\
\text{s.t. }&{\mathcal{E}_j(\mathcal{H}_{N^J}(\mathbf{x}_{design}))}^{\mathcal{F}_i}> 0\\
& lb \leq \mathbf{x}_{design} \leq ub\\
\forall &j \in \{1, ..., m\}\nonumber\\
\forall &i \in \{1, 2\}\nonumber
\end{align}
Where there are $m$ eigenvalues per form. Thus, all the eigenvalue constraints are stacked together. All the design variables also have box-constraints over them which are needed for modeling physically correct ranges for spring parameters. In our case, we allow the spring rest-lengths to vary between 50\% of the initial rest-length of the spring. $w$ is set to 0.001 for all examples described in the results section (Section \ref{sec:metaresults}).
\\
The above is a nonlinear optimization problem with nonlinear and box-constraints. We employ the Augmented Lagrangian method (ALM), \cite{Wright:1999:NO} to solve the same. A good refresher for the ALM method is also available in \cite{Narain:2012:AAR}. We use the standard ALM method and BFGS line search strategy in the inner loop. The above optimization also requires gradients of the energy and the Jacobian of the nonlinear constraints. We use finite-difference method for calculating these quantities.
%
%
\subsection{New spring addition -- modal analysis}\label{sec:modal}
Given a structure with a given spring configuration, modal analysis (\cite{Kry:2009:Modal}, \cite{Bharaj:2015:GV}) is a tool which can help calculate the deformation modes (via Eigenvalue analysis) of the structure. These modes (deformations) are the most likely changes in the structure as a result of excitation. A mode with the smallest non-positive eigenvalue given by equation \ref{eq:so} is the most likely to deform. If we were to run a forward simulation for a given shape with given spring configuration and rigid-body constraints, we would visually see such a deformation. Based of this observations we propose the following spring addition strategy.
\paragraph{Method} For all forms we perform eigenvalue decomposition to calculate eigenvalues ${\mathcal{E}_j(\mathcal{H}_{N^J}(\mathbf{x}_{design}))}^{\mathcal{F}_i}$ and corresponding eigenvalues ${\mathbf{e}_j(\mathcal{H}_{N^J}(\mathbf{x}_{design}))}^{\mathcal{F}_i}$. Then the largest non-positive eigenvalue/vector pair is selected. Intuitively, the eigenvector represents the velocities (linear and angular) in the null-space of the constraint Jacobian. Therefore, we back-project $\mathbf{e}_j^{\mathcal{F}_i}$ by the following operation \(\mathcal{E}_j\times(N_{J^T}\;\mathbf{e}_j^{\mathcal{F}_i})\) to calculate the velocities. Here, the unprojected velocity \(N_{J^T}\;\mathbf{e}_j^{\mathcal{F}_i}\) is multiplied with the corresponding \(\mathcal{E}_j\) to factor the intensity of negativity of the selected eigenmode.
\\
Using these velocities the rigid-body system is forward simulated by a single time-step (using Symplectic Euler) to calculate the positional and rotational deformations of the rigid-bodies. Finally, we calculate the deformation of the various vertices on the structure and select the corresponding rigid bodies (those vertices which deform the most) as candidate bars. These bars are then used to add a spring according the formulation proposed below.
\subsection{Minimal Energy Springs} \label{sec:minengsp}
A spring which has the same rest length on both forms will not increase the system potential, as a spring at rest-length does not add extra energy to the system. On form $\mathcal{F}_i$, a point on one of the bars is given by:
\(\mathbf{p} = t\mathbf{p}_1 + (1-t)\mathbf{p}_2\), where \(t\) is the
linear interpolation operator and \(\{p_1, p_2\}\) are ends of the bar. Similarly, candidate point on another bar is given by \(\mathbf{p}_c = s\mathbf{p}_{c1} + (1-s)\mathbf{p}_{c2}\). A distance metric between the two points \(f_j \in \mathcal{F}_i\) is \(d_1 = |\mathbf{p_c} - \mathbf{p}_1 |_2^2\). Then for \(f_j\) on $\{\mathcal{F}_i\}$, we want to minimize the following energy:
\begin{align}
\arg\min_{\{s, t\}} &\frac{1}{2} (d_i - d_k)^2\\
            \text{s.t. } & 0\leq s \leq 1 \label{eq:bx1}\\
                     & 0\leq t \leq 1\label{eq:bx2}\\
                     & i, k \in \{1, 2\}\nonumber
\end{align}
Here same \(\{s, t\} \in \mathcal{R}^1\) are used for both forms. By controlling the limits \(s, t\) we can avoid adding duplicates for contiguous bars on \(h_i\). Thus by adding box-constraints in equations \ref{eq:bx1} and \ref{eq:bx2}, which are greater than zero and less than one, consistent springs positions can be calculated. We employ BFGS with box-constraints to the solve the above optimization.
%
\subsection{Fabrication reformulation -- $z$-depth arrangement}
\label{sec:metafabrefor}
In the current rigid body formulation and optimization, self-intersection is not modeled, as a result, the springs are added in the same plane (assuming that all spring and bars are in the $x-y$ plane, with $z=0$). This method works for simulation and optimization, but will lead to sever self-intersections in the fabricated design. To avoid this situation much like \cite{Coros:2013:CDM,Bharaj:2015:CDW}, we add $z$-depth to each bar and spring such that, all each bar and spring has an unique $z$-depth. As a result, self-intersections are avoided and springs and bars of a structure can move freely and switch forms.
%
\begin{figure}[!t]
\centering
\includegraphics[width=\linewidth]{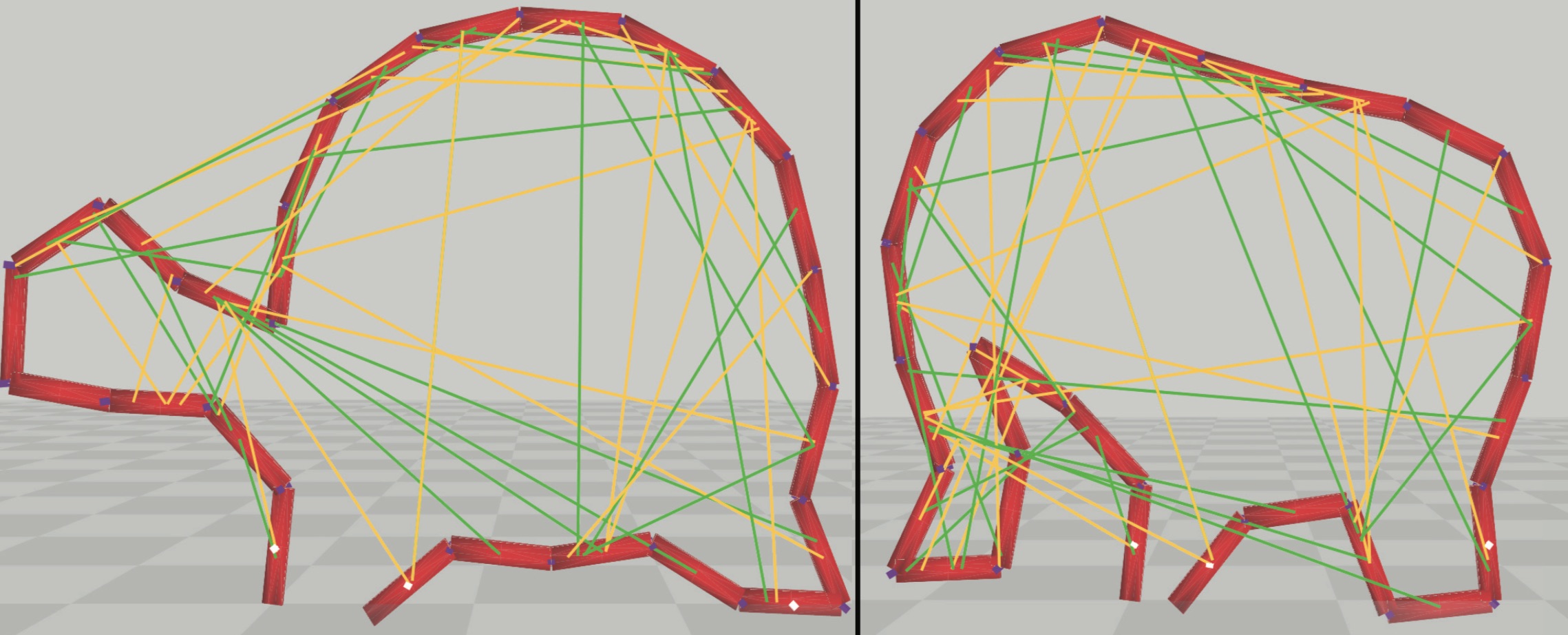}
\caption[Higher Complexity]{\textbf{Higher Complexity}: Turtle (Left) and Elephant (Right) are used to create complex forms for a structure.}
\label{fig:highcomplex}
\end{figure}
\section{Fabrication and results}\label{sec:metaresults}
This section discusses the various metamorphs we optimized and fabricated. In-order to fabricate the spring model used for the numerical simulation, we created a spring assembly process. The fabricated springs match the deformation behavior and potential energy properties used in numerical simulation. We first discuss the details for the spring assembly and then present optimized results with virtual and fabricated validation, timings, and implementation details.
%
\subsection{Fabrication and Calibration}\label{sec:metafabrication}
\begin{figure*}[t!]
    \centering
    \begin{subfigure}[t]{0.5\textwidth}
      \centering
      \includegraphics[width=\linewidth]{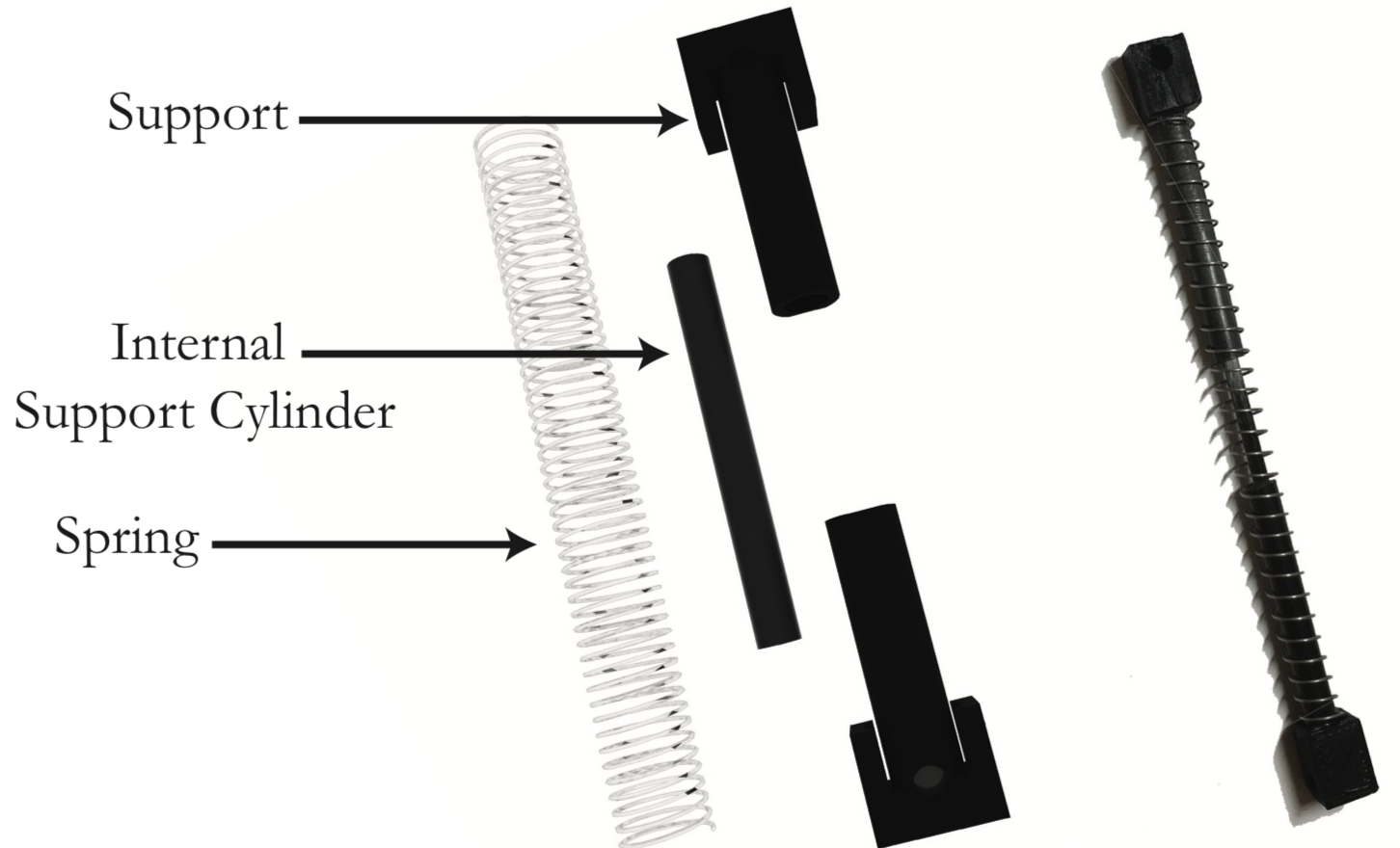}
      \caption[Spring assembly]{The diagram on the left shows the parts used to assemble and create a fabricated spring shown in the figure on the right}
      \label{fig:springassembly}
    \end{subfigure}%
    ~\hspace{10px}
    \begin{subfigure}[t]{0.5\textwidth}
      \centering
      \includegraphics[width=\linewidth]{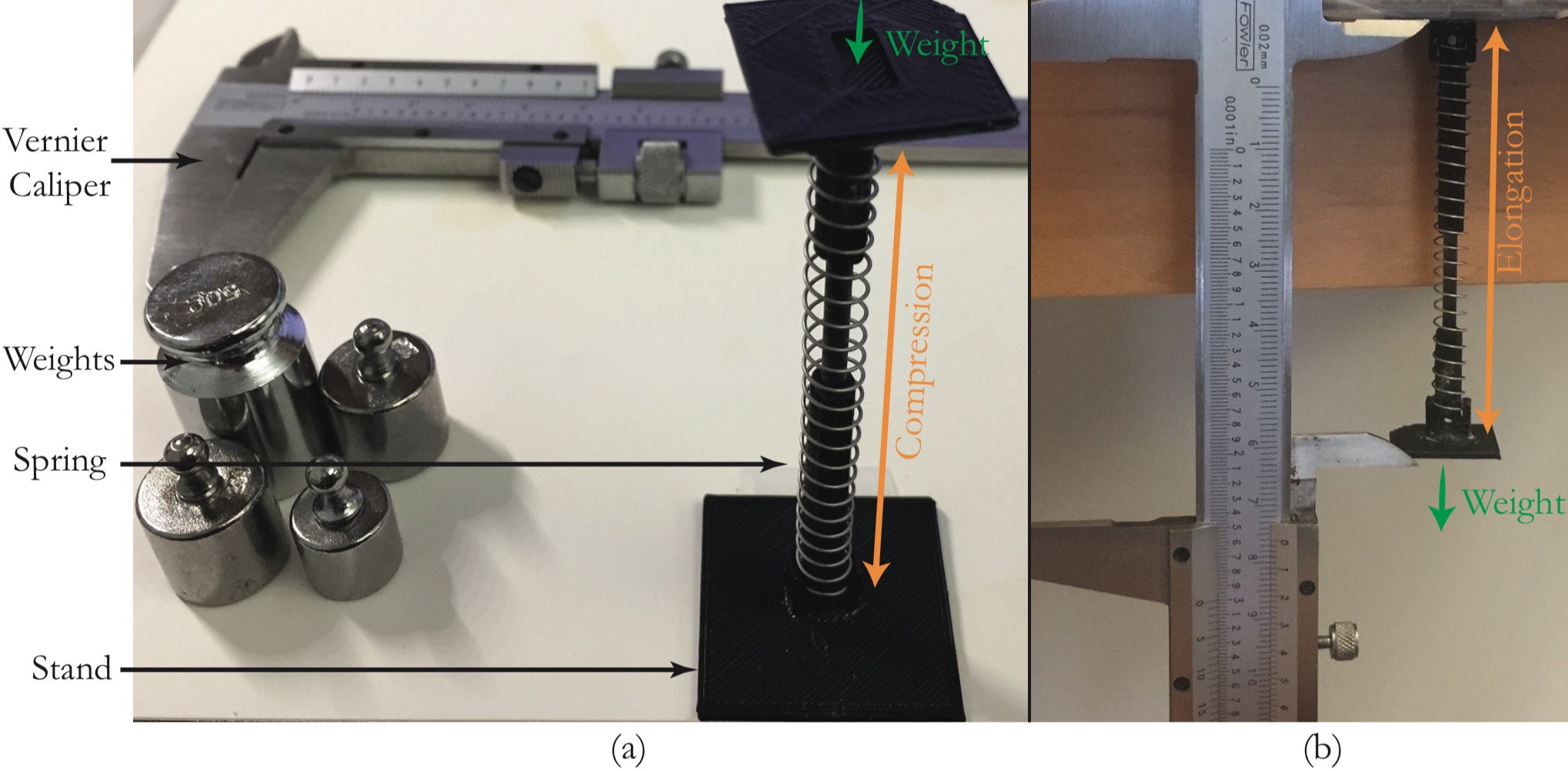}
      \caption[Spring calibration]{(a) Compression, and (b) elongation measurement setup. Various weights are either rested on top or hung-on to measure compression and elongation, respectively of the spring.}
      \label{fig:springCalibration}
    \end{subfigure}
    \caption{(a) Spring assembly, and (b) Spring calibration}
\end{figure*}
\paragraph{Fabrication methodology}
The basic requirements of a spring are that the compression and elongation of the spring should happen in a straight line (in 3D). If we have a simple steel-wire spring eventually it will bend without internal support. However, the internal support should not lead to change in spring's stiffness or rest length properties. With these requirements in mind, we create an assembly process shown in Figure \ref{fig:springassembly} to create a spring. Each spring consist of: a steel wire spring, an internal support (3D printed support with a hollow cylinder), and an internal support cylinder made of carbon-fiber. The cylinders are light weight and have very low coefficient of friction, as a results can slide very easily into the internal support and more importantly do not increases or decrease the spring's stiffness. Finally, all parts are put together, and the ends of the spring are super-glued to the ends of the internal supports as shown in Figure \ref{fig:springassembly} (right). This assembly process leads to springs which move in a straight line in any 3D orientation.
\paragraph{Spring calibration}
Our spring have two properties, stiffness ($k$) and rest length. In-order to have simple near linear spring stiffness, we choose McMaster-Carr's Corrosion-Resistant Compression Spring Stock with 0.25" OD, 0.216" ID. This spring has a near linear spring stiffness. We then use the setup shown in Figure \ref{fig:springCalibration} to measure the actual spring stiffness. As shown, a stand is created to hold the spring vertically in place, and various weights are rested on top of the spring or hung from it. Then vernier calipers are used to measure the compression/elongation in the spring for the said weight. We use Hook's spring formula $F=kX$, where $X$ is the amount by which the free end of the spring gets displaced from its rest length, $F=mg$ is the force acting on the spring, $m$ is the mass of the weight and $g$ is the acceleration due to gravity. For each spring and for each weight, we measure three times for $X$, and then use the average $X$ (over all measurements). The correct $k$ for the spring (for a given weight) is calculated accordingly. Since our springs are linear, we get the nearly the same $k$ for each weight for compression and elongation.
%
%
\subsection{Results}
\paragraph{Generic metamorphs}
%
%
Figure \ref{fig:metateaser} shows an example of structure that can morph from a duck-like into teddy-like structural form. This structure consists of four bars that are connected by hinge connections (purple) and held at end-points (white). The iterative scheme optimizes for gravitational external forces so that both forms of the structure are statically stable. The optimized forms are then fabricated using the methodology described above and is shown in the corresponding figure with black bars. We also create a more complex form of the duck and teddy example as shown in Figure \ref{fig:duckTeddy}, this example shows that the we can scale-up in complexity for a given Metamorph. 
\\
Figure \ref{fig:highcomplex} shows our most complex example, where two drastically different input curves, turtle and elephant are optimized for bistability. For all the above examples we provide virtual validation by running a forward rigid body dynamics simulation for both optimized stable forms. While for Figure \ref{fig:metateaser} we show fabricated validation results. The convergence timings and complexity for each example are shown in Table \ref{tbl:complexity}.
%
\begin{figure*}[!t]
\centering
\includegraphics[width=\linewidth]{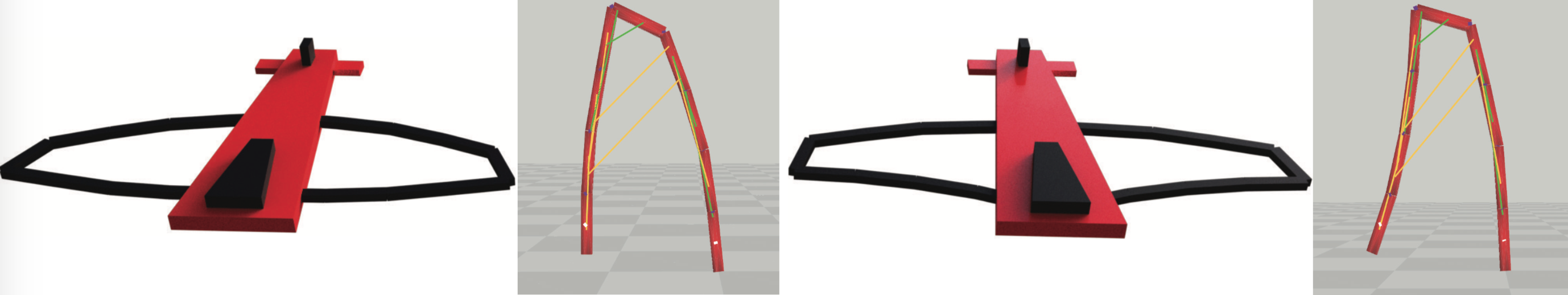}
\caption[Functional Metamorph]{\textbf{Bistable Wing}: Concept plane design with two stable wing forms, and the corresponding spring based structural forms}
\label{fig:functional}
\end{figure*}
\paragraph{Functional metamorphs} A natural consequence of bistability is that the same structure can change form. As a result it becomes useful for multiple use scenarios. For example, \cite{Chen14:ANM} shows an example of a cloth hanger and a phone holder, where a single form is optimized for. We show a use case inspired from recent work in the field of robotics for bistable wings \cite{manchester2017variable}, where a linkage-based structure was used for a change in the wing orientation. Our method is flexible enough that we can not only change the orientation of the wing, but also its form. Figure \ref{fig:functional} show the concept of a functional wing that can take two different forms and change the amount of air-drag acting on the wing. Such a bistable wing is useful in different scenarios such as, plane landing, perching, or cruising. By combining our approach with \cite{Umetani2014} we can create two wing designs for a single plane!
%
%
\paragraph{Posable metamorphs} 
Creating virtual character with deformable articulate forms is now possible \cite{Martin:2011:EEM}. There has been a push to achieve the same for fabriable characters such as \cite{Skouras:2013:CDA:2461912.2461979}. In such works although deformations are quite pronounced, the deformed forms are not stable in the second-order sense (Section~\ref{sec:metaoverview}) and need constant external forces to hold the forms.

\begin{figure}[!t]
    \includegraphics[width=\linewidth]{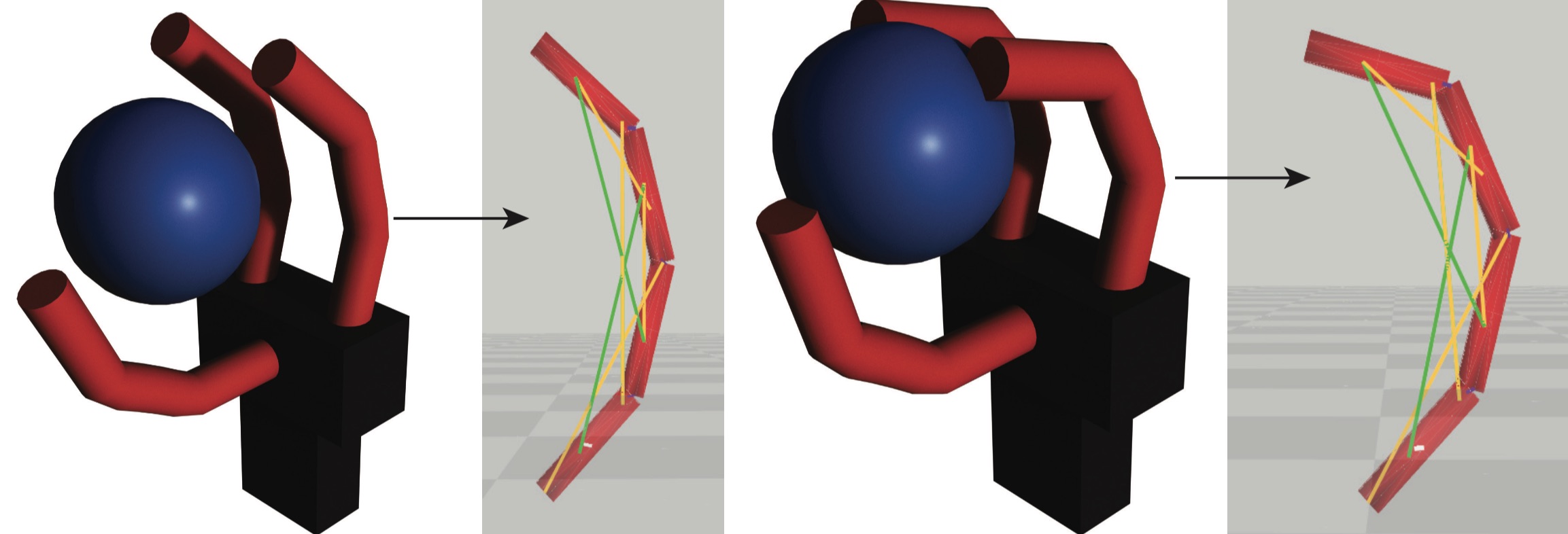}
    \caption[Posable Hand/Gripper]{\textbf{Posable Hand:} Selected key-frames of character's hand are used to create the hand model. Different fingers have different stable forms. This can be used for stop-motion character animation and as a gripper.}
    \label{fig:posable}
\end{figure}

We create a example-based poseable \emph{hand} (Figure \ref{fig:posable}) that does not have these limitations. This can not only be used as a gripper, but also with puppets and paper mache characters. The example shown consists of two finger and a \emph{thumb}, where each finger is designed to have different stable forms. Such a setup can also be used for holding non-convex shapes, where the desired grip (finger forms) are input to the optimization method.
%
\paragraph{Implementation}
The rigid body, spring energy and optimization frameworks were written in C++ and run on Intel Quadcore CPU, on a single thread. \textit{Alglib} a C++ library was used for Augmented Lagrangaian Method and BFGS. \cite{press1996numerical}'s method based on QR decomposition was used for Eigen value decomposition. Matlab's \textit{syms} package was used for calculating and testing against analytically calculated gradient and Jacobian of spring's potential energy. 

Table \ref{tbl:complexity} shows the complexity -- number of bars, and springs, and optimization convergence timings for each example discussed above. While the fabricated example shown in Figure \ref{fig:metateaser} took about a day to fabricate and assemble.
\begin{table}[]
\centering
\caption[Metamorphs: Space and time complexity]{Space and time complexity of various examples}
\label{tbl:complexity}
\small
\begin{tabular}{lccc}
\hline
Metamorph             &  No. Bars & No. Springs & Timing (mins.) \\ \hline
Duck teddy, low-res.  &         6            &     4               &         0.08                     \\
Duck teddy, high-res. &         14           &     21              &         13.16                      \\
Turtle elephant       &         24           &     47              &         21.15                     \\
Plane wings           &         9            &     14              &         0.48                      \\
Hand finger           &         4            &     6               &         0.02                      \\ \hline
\end{tabular}
\end{table}
\section{Limitations and Future Work}
\label{sec:future}
We propose a novel computational design tool for creating bistable planar structures with second-order guarantees on the stability of each form. The iterative optimization uses modal analysis to first choose a location to add an internal support spring, then a nonlinear optimization is used to optimize for first and second-order energy stability. Such structures can have drastically different forms.
While the current method is promising, there are limitations of the method that can lead to future research. 
\\
Because the input forms (curves) for a structure can be non-convex, newly added internal support springs may not stay inside the structure during the optimization. Because the algorithm tries to find a minimal energy springs (Section \ref{sec:minengsp}), it can result in springs remaining out of the convex form. In Figure \ref{fig:duckTeddy}, the springs near the beak of the duck remain outside the form. Another limitation of the method is that it solves for planar structures only. Although, switching to 3D springs would be easy we limit ourselves to planar structures due to fabrication constraint. That is, each spring must be free to move in a $z$-plane and avoid self-intersections. Note that this also true for all general linkages. We experimented with ball-and-socket joints, but again due to self-intersections chose to use planar hinge joints only.
\\
As shown in Section \ref{sec:metaresults}, we add about $n$ springs for a $n$ degrees-of-freedom structural forms. Each spring is about $0.5$ cms thick when fabricated, as a result if one were to fabricate such a structure, we'd go $0.5n$ cms deep along the $z$-axis. This can lead to extra torques/forces along the $z$-axis and is aesthetically displeasing. Hence, newer ways of fabrication are needed that can lead to thinner internal support springs with reduced $z$-depth. And, facilitate with building complex bistable structures.
\section{Appendix}
\label{ch:app}
\subsection{Rigid Body Dynamics - Physical Simulation}
\label{sec:simulation}

Each automata is modeled as a rigid multi-body system. Since the mechanisms we optimize typically exhibit numerous kinematic loops, we opt for a maximal coordinates dynamics formulation. Therefore, the state of each rigid body $i$ consists of position and orientation degrees of freedom $\bq_i$, and their linear and angular velocity derivatives $\dot{\bq}_i$. The vectors $\bq$ and $\dot{\bq}$ concatenate the states of all rigid bodies in the system.

We model joints, virtual motors, and frictional contacts using a set of constraints of the form $\bC(\bq) = \mathbf{0}$, and their time derivatives $\dot{\bC}(\bq) = \dot{\bC}^d$ \cite{Cline03}. According to the principle of {\em virtual work}, the constraints give rise to internal forces $\mathbf{f}_c = \bJ^T\mathbf{\lambda}$, where $\bJ$ denotes the Jacobian $\frac{\partial{C}}{\partial{\bq}}$, and $\mathbf{\lambda}$ are Lagrange multipliers that intuitively correspond to the magnitudes of the generalized forces needed to satisfy each constraint. To integrate the motion of the mechanisms forward in time, we must first compute the constraint forces $\mathbf{f}_c$. Without loss of generality, we can express their magnitudes implicitly as:
\begin{equation}
\label{eq:lambda}
\mathbf{\lambda} = -k_p \bC(\bq_{t+1}) - k_d (\dot{\bC}(\bq_{t+1}) - \dot{\bC}^d)
\end{equation}
where subscript $t$ indicates the time instance, and the coefficients $k_p$ and $k_d$ allow us to set the relative stiffness of different types of constraints. A Taylor-series approximation of the position constraints allows us to express $\bC(\bq_{t+1})$ as:
\begin{equation}
C(\bq_t+h\dot{\bq}_{t+1}) \dot{=} \bC(\bq_t) + h J^T \dot{\bq}_{t+1}
\end{equation}
where $h$ denotes the time step. Using the chain rule, the time-derivative of the constraints can be written as $\dot{\bC}(\bq_{t+1}) = \bJ^T \dot{\bq}_{t+1}$. This allows us to approximate Eq. \ref{eq:lambda} as:
\begin{equation}
\label{eq:jqDot}
\bJ \dot{\bq}_{t+1} = -a \mathbf{\lambda} - a k_p \bC(\bq_t) + k_d a \dot{\bC}^d
\end{equation}
where $a = \frac{1}{h k_p + k_d}$. Using the equations of motion of the multi-body system, the generalized velocities $\dot{\bq}_{t+1}$ are given by:
\begin{equation}
\label{eq:qDot}
\dot{\bq}_{t+1} = \dot{\bq}_t + h \bM^{-1}(\bF_{ext} + \bJ^T\mathbf{\lambda})
\end{equation}
where $\bM$ denotes the system's mass matrix, and the term $\bF_{ext}$ stores the gravitational forces acting on the system. Multiplying Eq.~\ref{eq:qDot} by $\bJ$, and combining the result with Eq.~\ref{eq:jqDot}, results in the following system of equations that is linear in $\mathbf{\lambda}$:
\begin{equation}
\label{eq:lambdaLinSystem}
\bA \mathbf{\lambda} = \mathbf{b}
\end{equation}
where $\bA = h \bJ \bM^{-1} \bJ^T + a\bI$ and $\mathbf{b} = k_d a \dot{\bC}^d - a k_p \bC(\bq_t) - \bJ\dot{\bq}_t - h \bJ \bM^{-1}\bF_{ext}$. Because the constraint forces arising from frictional contacts are subject to inequality constraints, as discussed shortly, rather than solving Eq.~\ref{eq:lambdaLinSystem} directly, we follow the work of Smith et al.~\cite{SKVTG2012} and compute $\mathbf{\lambda}$ by solving a quadratic program:
\begin{equation}
\label{eq:lambdaQPSystem}
\min_\lambda \frac{1}{2} (\bA \mathbf{\lambda} - \mathbf{b})^T (\bA \mathbf{\lambda} - \mathbf{b}) s.t. \mathbf{D\lambda} \geq \mathbf{0}
\end{equation}
where the matrix $\mathbf{D}$ stores all the inequality constraints that need to be enforced. Once the constraint forces are computed, we use Eq.~\ref{eq:qDot} to compute the generalized velocity term $\dot{\bq}_{t+1}$, and the positional degrees of freedom $\bq_{t+1}$ are integrated forward in time as described by Witkin \cite{witkin2001}.

The derivation we provide here is related to methods implemented by some modern rigid body engines, such as the Open Dynamics Engine \cite{ode:2008}. However, rather than being restricted to working with ad-hoc parameters that hold little physical meaning, such as the Constraint Force Mixing term, Error Reduction Parameter and the Parameter Fudge Factor, we control the behavior of our simulations by manipulating the stiffness and damping parameters, $k_p$ and $k_d$, which are set independently for each constraint type (as detailed below). In the limit, as $k_p$ goes to infinity and $k_d$ to $0$ (i.e., infinitely stiff spring), this formulation remains well-defined, and corresponds to solving the constraints exactly. However, from the point of view of numerical stability, it is often better to treat the constraints as stiff implicit penalty terms.
\paragraph{Pin joints} that allow a pair of components to rotate relative to each other about a pre-specified axis are implemented using two sets of constraints. First, we ensure that the coordinates of the pin coincide in world space using a vector-valued constraint of the form $\mathbf{C}(\bq)=\bx(\bq_i(t),\bp_i)-\bx(\bq_j(t),\bp_j)$. Here, $\bx(\bq_a,\bp)=\bt_a + \bR_a\bp$ corresponds to the world coordinates of the point $\bp$, $\bt_a\in\mathbb{R}^3$ is defined as the position of center of mass of rigid body $a$, and $\bR_a$ corresponds to its orientation. The location of the pin joint is defined by specifying the local coordinates of the pin, $\bp_i$ and $\bp_j$, in the coordinate frames of the two rigid bodies $i$ and $j$ that are connected to each other. To ensure that the two rigid bodies rotate relative to each other only about the pre-scribed axis, we use an additional vector-valued constraint, $\mathbf{C}(q)=\bR_i \bn_i - \bR_j \bn_j$, where $\bn_i$ and $\bn_j$ represent the coordinates of the rotation axis in the local coordinates of the two rigid bodies, and are set to $(0,0,1)^T$ for all our experiments. The $k_p$ and $k_d$ coefficients for the pin joint constraints are set to $10^8$ and $10^4$, respectively.

\paragraph{Motor constraints} are used to mimic the effect of physical actuators. For this purpose, we prescribe the time-varying, desired relative angle between a select set of rigid body pairs. In particular, we assume that each limb of the mechanical toys has an input crank that operates relative to the main body. As we already employ pin joint constraints between these pairs of rigid bodies, the motor constraints directly measure the difference between their relative orientation and the target motor angle. The target motor angles are specified by {\em phase profile functions} $f(\alpha)$, as described by Coros et al. \cite{Coros:2013:CDM}. The desired value for the time derivative of the constraint, $\dot{C}^d$, is set to $\dot{f}(\alpha)$, and it intuitively corresponds to the target velocity of the virtual motor. The $k_p$ and $k_d$ coefficients for the motor constraints are set to $10^8$ and $10^5$, respectively. 

\paragraph{Frictional contacts} move our automata around their simulated environments, and friction and contact forces must be bounded to generate physically-plausible results. Each contact introduces three constraints. Let $\bn$ denote the contact normal. The first constraint specifies that the penetration distance, measured along the normal, should be $0$: $C(\bq_a) = \bn^T(\bx(\bq_a,\bp)-\bx_p)$. Here, $\bp_a$ corresponds to the coordinates of the contact point in the frame of rigid body $a$, and $\bx_p$ is the projection of the contact point onto the environment. For this constraint, $k_p=10^8$, $k_d=10^4$, and, importantly, the constraint force magnitude is constrained to be positive: $\lambda_n \geq 0$.

To model friction, we employ a pyramid approximation to the friction cone, as is standard in real-time simulation systems. More precisely, we let $\bt_1$ and $\bt_2$ be two orthogonal vectors that are tangent to the contact plane, and define constraints similar to the one for the normal direction, but acting along the tangent vectors. However, friction forces should only act to reduce the relative velocity at the contact point to $0$. For this reason, we set $k_p$ to 0 for these constraints, while $k_d$ is set to $10^4$. To ensure that tangential forces remain within the friction pyramid, we add inequality constraints of the form $-\mu \lambda_n \leq \lambda_t \leq \mu \lambda_n$ for the magnitude of the tangential forces acting along $\bt_1$ and $\bt_2$, where $\mu$ represents the friction coefficient.
%
\subsection{Axis-Angle representation}
\label{sec:axisangle}
A circular movement of angle $\theta$ around a specified axis $\mathbf{\bar{v}}$ in $\mathbb{R}^3$ is given by axis-angle:
\begin{align}
\mathbf{v} &= \theta \mathbf{\bar{v}}\\
\theta &= ||\mathbf{v}||\\
\mathbf{\bar{v}} &= \frac{\mathbf{v}}{||\mathbf{v}||}
\end{align}
The rotation-matrix (from the axis-angle) is given by Euler-Rodrigues's exponential coordinates (\cite{Murray:1994:AMIRM}, Page 29):
\begin{align}
\mathrm{R} &= \mathrm{I} + sin(\theta)[\mathbf{\bar{v}}]_{\times} + (1-cos(\theta))[\mathbf{\bar{v}}]^2_{\times}
\end{align}
$\mathbf{\bar{v}}$ is a unit-vector, so,
\begin{align}
[\mathbf{\bar{v}}]^2_{\times} &= \mathbf{\bar{v}}\mathbf{\bar{v}}^T - \mathrm{I}\\
\mathrm{R} &= cos(\theta)\;\mathrm{I} + sin(\theta)[\mathbf{\bar{v}}]_{\times} + (1-cos(\theta))\mathbf{\bar{v}}\mathbf{\bar{v}}^T
\end{align}
Also, $[\mathbf{a}]_{\times}$ is a skew-symmetric matrix:
\begin{align}
[\mathbf{a}]_{\times} = \left( \begin{array}{ccc}
0 & -a_3 & a_2 \\
a_3 & 0 & -a_1 \\
-a_2 & a_1 & 0 \end{array} \right) \in \text{Skew}_3
\end{align}
\paragraph{Axis-Angle --- Gradient}
Let $\mathbf{u}' = \mathrm{R(\mathbf{v})}\,\mathbf{u}$, then we need to caluclate $\frac{\partial \mathbf{u}'}{\partial v_i}$. As $\mathbf{u}$ is independent of $\mathbf{v}$. We get the following (derivation in \cite{Gallego:2015:CFD}, Appendix E):
{\small
\begin{align}
  \frac{\partial \mathbf{u}'}{\partial v_i} = \frac{\partial \mathrm{R}(\mathbf{v})}{\partial v_i}\mathbf{u}\\
   \frac{\partial \mathrm{R}}{\partial v_i} = cos(\theta)\bar{v_i}[\mathbf{\bar{v}}]_{\times} + sin(\theta)\bar{v_i}[\mathbf{\bar{v}}]^2_{\times} + \frac{sin(\theta)}{\theta}[\mathbf{e}_i - \bar{v_i} \mathbf{\bar{v}}]_{\times} +\nonumber\\
   \frac{1-cos(\theta)}{\theta}(\mathbf{e}_i \mathbf{\bar{v}}^T - \mathbf{\bar{v}}\mathbf{e}_i^T - 2\bar{v_i}\mathbf{\bar{v}}\mathbf{\bar{v}}^T)\label{eq:dRdv}
\end{align}
}
Note that, $\frac{\partial \mathbf{u}'}{\partial v_i}$ is a [$3\times 1$] column vector for $\mathbf{v} = \{v_1, v_2, v_3\}^T$. More compact gradient is given by, for example, \cite{Gallego:2015:CFD}.

\subsection{Spring --- Potential, Gradients and Hessians}
Given a rigid-bodies $\mathcal{B}_i$, the world position $\mathbf{p}_i$ of local point $\mathbf{u}_i$ is given by:
\begin{align}
  \mathbf{p}_i( \mathbf{c}_i, \mathbf{v}_i) = \mathrm{R}_i(\mathbf{v}_i) \mathbf{u}_i + \mathbf{c}_i
\end{align}
Then, the spring potential between local points $\mathbf{u}_1, \mathbf{u}_2$ on $\mathcal{B}_1, \mathcal{B}_2$ respectively is given by:
\begin{align}
  \mathrm{V}(\mathbf{x}) &= \frac{1}{2} k (l - \sqrt{f(\mathbf{x})}\;)^2\\
  \mathbf{x} &= \{\mathbf{c}_1, \mathbf{v}_1, \mathbf{c}_2, \mathbf{v}_2\}\\
  f(\mathbf{x}) &= g(\mathbf{x})^T\,g(\mathbf{x})\label{eq:DP}\\
  g(\mathbf{x}) &= [\mathrm{R}_1(\mathbf{v}_1) \mathbf{u}_1 + \mathbf{c}_1 - \mathrm{R}_2(\mathbf{v}_2) \mathbf{u}_2 - \mathbf{c}_2]
\end{align}
For vector-spaces, we have the following property:
\begin{align}
  \dfrac{\mathrm d}{\mathrm dx}\left({ \mathbf{r}\left({x}\right) \cdot \mathbf{r}\left({x}\right) }\right) &= \mathbf{r}\,'\left({x}\right)\cdot\mathbf{r}\left({x}\right) + \mathbf{r}\left({x}\right)\cdot\mathbf{r}\,'\left({x}\right)\\
   &\equiv 2 \; \mathbf{r}\left({x}\right)\cdot\mathbf{r}\,'\left({x}\right)\label{eq:gradDP}
\end{align}

\paragraph{Spring --- Gradient}\label{sec:gradient}
\begin{align}
    \frac{\partial \mathrm{V}(\mathbf{x}) }{\partial x_i} = -\frac{k}{2}\frac{(l - \sqrt{f(\mathbf{x})}\;)}{\sqrt{f(\mathbf{x})}} \frac{\partial f(\mathbf{x})}{\partial x_i}\\
    \text{Using \ref{eq:DP} and \ref{eq:gradDP}, for}\;j=\{1, 2, 3\}\; \nonumber\\
    \frac{\partial \mathrm{V}(\mathbf{x}) }{\partial x_i} = \frac{k}{2}\Bigg(1 - \frac{l}{\sqrt{g(\mathbf{x})^T\,g(\mathbf{x})}}\Bigg) \frac{\partial g(\mathbf{x})^T\,g(\mathbf{x})}{\partial x_i}\\
    \equiv k\Bigg(1 - \frac{l}{\sqrt{g(\mathbf{x})^T\,g(\mathbf{x})}}\Bigg)\Bigg(g(\mathbf{x})^T \frac{\partial \,g(\mathbf{x})}{\partial x_i}\Bigg)\label{eq:gradV}\\
\frac{\partial \mathrm{V}(\mathbf{x}) }{\partial c_{1j}} = k\Bigg(1 - \frac{l}{\sqrt{g(\mathbf{x})^T\,g(\mathbf{x})}}\Bigg)(g(\mathbf{x})^T \mathbf{e}_j)\\
\frac{\partial \mathrm{V}(\mathbf{x}) }{\partial c_{2j}} = -k\Bigg(1 - \frac{l}{\sqrt{g(\mathbf{x})^T\,g(\mathbf{x})}}\Bigg)(g(\mathbf{x})^T \mathbf{e}_j)\\
\text{Using \ref{eq:dRdv}},\nonumber\\
\frac{\partial \mathrm{V}(\mathbf{x}) }{\partial v_{1j}} = k\Bigg(1 - \frac{l}{\sqrt{g(\mathbf{x})^T\,g(\mathbf{x})}}\Bigg)\Bigg(g(\mathbf{x})^T \frac{\partial \mathrm{R_1(\mathbf{v_1})}\mathbf{u_1}}{\partial v_{1j}}\Bigg)\\
\frac{\partial \mathrm{V}(\mathbf{x}) }{\partial v_{2j}} = -k\Bigg(1 - \frac{l}{\sqrt{g(\mathbf{x})^T\,g(\mathbf{x})}}\Bigg)\Bigg(g(\mathbf{x})^T \frac{\partial \mathrm{R_2(\mathbf{v_2})}\mathbf{u_2}}{\partial v_{2j}}\Bigg)
\end{align}
\paragraph{Spring --- Hessian}\label{sec:hessian}
Let $h(\mathbf{x})_i = g(\mathbf{x})^T \frac{\partial \,g(\mathbf{x})}{\partial x_i}$. The hessian is a $12\times12$ square-matrix, with $\{\mathbf{x}\}^{12\times1}$.  Using equation \ref{eq:gradV}, we get:
\begin{align}
\frac{\partial^2}{\partial x_j}\frac{\mathrm{V}(\mathbf{x}) }{\partial x_i} = k\frac{\partial}{\partial x_j}\Bigg(\Bigg(1 - \frac{l}{\sqrt{g(\mathbf{x})^T\,g(\mathbf{x})}}\Bigg)\Bigg(g(\mathbf{x})^T \frac{\partial \,g(\mathbf{x})}{\partial x_i}\Bigg) \Bigg)\\
= k\Bigg( \frac{\partial}{\partial x_j} \Bigg(g(\mathbf{x})^T \frac{\partial \,g(\mathbf{x})}{\partial x_i}\Bigg) - l \frac{\partial}{\partial x_j} \Bigg(\frac{g(\mathbf{x})^T \frac{\partial \,g(\mathbf{x})}{\partial x_i}}{\sqrt{g(\mathbf{x})^T\,g(\mathbf{x})}} \Bigg)\Bigg)\\
= k\Bigg( \frac{\partial}{\partial x_j} h(\mathbf{x})_i - l \frac{\partial}{\partial x_j} \frac{h(\mathbf{x})_i}{\sqrt{g(\mathbf{x})^Tg(\mathbf{x})}}\Bigg)\\
\text{Using \ref{eq:gradDP}},\nonumber\\
\frac{\partial}{\partial x_j} h(\mathbf{x})_i = \frac{\partial g(\mathbf{x})}{\partial x_j}^T\frac{\partial g(\mathbf{x})}{\partial x_i} + g(\mathbf{x})^T\frac{\partial^2 g(\mathbf{x})}{\partial x_j\, x_i}\label{eq:d2gdxij}\\
\text{Using chain-rule:}\nonumber\\
\frac{\partial}{\partial x_j} \frac{h(x)_i}{\sqrt{g(\mathbf{x})^Tg(\mathbf{x})}} = -\frac{h(\mathbf{x})_j h(\mathbf{x})_i}{2(g(\mathbf{x})^Tg(\mathbf{x}))^{\frac{3}{2}}} + \frac{\frac{\partial}{\partial x_j} h(\mathbf{x})_i}{\sqrt{g(\mathbf{x})^Tg(\mathbf{x})}}
\end{align}

Now, we need to define the $\frac{\partial^2 g(\mathbf{x})}{\partial x_j\, x_i}$ term in equation \ref{eq:d2gdxij}, rest are defined below:
\begin{align}
  \frac{\partial g(\mathbf{x})}{\partial c_{1i}} &= \mathbf{e}_i\\
  \frac{\partial g(\mathbf{x})}{\partial v_{1i}} &= \frac{\partial \mathrm{R}_1(\mathbf{v}_1)}{\partial v_{1i}}\mathbf{u}_1\\
  \frac{\partial g(\mathbf{x})}{\partial c_{2i}} &= -\mathbf{e}_i\\
  \frac{\partial g(\mathbf{x})}{\partial v_{2i}} &= -\frac{\partial \mathrm{R}_2(\mathbf{v}_2)}{\partial v_{2i}}\mathbf{u}_2
\end{align}
With $l, k= \{1, 2\}$, all terms of the form $\frac{\partial^2 g(\mathbf{x})}{\partial c_{lj} c_{ki}}$, $\frac{\partial^2 g(\mathbf{x})}{\partial v_{lj} c_{ki}}$, and $\frac{\partial^2 g(\mathbf{x})}{\partial v_{lj} v_{ki}}$, $l \neq k$ are $\mathbf{0}$. We now need to define the following:
\begin{align}
  \frac{\partial^2 g(\mathbf{x})}{\partial v_{1j} v_{1i}} &= \frac{\partial}{\partial v_{1j}} \frac{\partial \mathrm{R}_1(\mathbf{v}_1)}{\partial v_{1i}}\mathbf{u}_1\\
  \frac{\partial^2 g(\mathbf{x})}{\partial v_{2j} v_{2i}} &= \frac{\partial}{\partial v_{2j}} \frac{\partial \mathrm{R}_2(\mathbf{v}_2)}{\partial v_{2i}}\mathbf{u}_2
\end{align}
\bibliographystyle{ACM-Reference-Format}
\bibliography{lib}

\end{document}